# Thinging Machines for Requirements Engineering: Superseding Flowchart-Based Modeling

**Sabah Al-Fedaghi** *

*Computer Engineering Department*
*Kuwait University*
*Kuwait*
**salfedaghi@yahoo.com, sabah.alfedaghi@ku.edu.kw**

*Abstract* - **This paper directs attention to conceptual modeling approaches that integrate advancements and innovations in requirements engineering. In some current (2024) works, it is claimed that present elicitation of requirements models focus on collecting information using natural language, which yields ambiguous specifications. It is proposed that a solution to this problem involves using complexity theory, transdisciplinarity, multidimensionality and knowledge management. Examples are used to demonstrate how such an approach helps solve the problem of quality and reliability in requirements engineering. The modeling method includes flowchart-like diagrams that show the relationships among system components and values in various modes of operation as well as path graphs that represent the system behavior. This paper focuses on the diagrammatic techniques in such approaches, with special attention directed to flowcharting (e.g., UML activity diagrams, business process model and notation (BPMN) business process diagrams). We claim that diagramming methods based on flowcharts is an outdated technique, and we promote an alternative diagrammatic modeling methodology based on thinging machines (TMs). TMs involve a high-level diagrammatic representation of a real-world system that integrates various component specifications to be refined into a more concrete executable form. TM modeling is a valuable tool to integrate requirements elicitation and address present challenges comprehensively. To demonstrate that, case studies are re-modeled using TMs. A TM model involves static, dynamic diagrams and event chronology charts. This study contrasts the flowchart-based and the TM approaches. The results point to the benefits of adopting the TM diagramming method.**

## I. INTRODUCTION

Requirements engineering usually includes the elicitation, analysis, validation, and specification of requirements. The specification of a requirement must be expressed in language that everyone involved can clearly understand [1]. Uncertain requirements may result in the production of unsatisfactory software systems [2].

### *A. About Requirements*

The requirements elicitation constitutes the basis on which the other phases of the life cycle are carried out [3]. Requirements elicitation is the process of seeking, uncovering, acquiring, and elaborating requirements for software systems [4]. Typically, this requires designing a semiformal model to document the requirements using diagrams, textual descriptions, and formal elements (e.g., logic).

One option in this context is a conceptual model that represents user requirements by specifying what the system *does* using a level of abstraction that is close to the problem and to the stakeholders [5]. Conceptual modeling involves a high-level representation of a real-world system that integrates various components to refine it into a more concrete (computer) executable form. Several processes, norms, standards, and initiatives have been utilized in conceptual models.

### *B. Aim*

A central aim in this paper is to deliberate on the type of conceptual modeling approach that integrates advancements and innovations in requirements elicitation to address present challenges comprehensively. Without losing generality and for the sake of limiting the research materials, this paper focuses on two recent approaches.

First, [3] proposed integrating 'revolutionary theories' and concepts into requirements engineering, to structure and apply a model to manage the needs of users based on: (a) Complexity theory, because the world of software is complex; (b) Transdisciplinarity, because a broad vision is required to understand and solve complex problems; (c) Multidimensionality, because the systems are not one-dimensional; and (d) Knowledge management, because software engineers must adequately interpret the context of the problem and because it is an essential tool to model, understand and navigate the world, and to understand the problem to be solved. Consequently, [3] developed a model intended to "[help] to solve the problem of quality and reliability of requirements engineering."

Second, an alternative diagrammatic methodology called a thinging machine (TM) has been proposed as a base for conceptual modeling in requirement engineering. A TM model can be used as a blueprint to document various aspects of a system. Additionally, it is likely to be used (with further future refinements) to generate software code similar to that in the UML model approach, known as model-driven development.

----------------------

* Retired June 2021, seconded fall semester 2021/2022

Furthermore, the TM model is used as a draft to construct diagrams of parts of a system utilized for communication among developers and stakeholders.

Even though the paper is limited to these two approaches, the underlining claim is that TM modeling is a viable methodology in the attempts to integrate advancements and innovations in requirements elicitation to address present challenges comprehensively.

### C. *Problem*

According to [3], software development has not yet overcome the so-called software crisis of the 1960s, so the technological product continues to be delivered with deficiencies in quality, security, and reliability. Currently (2024), methods used for the requirements elicitation phase are informal and based on natural language, which makes the software engineering process more complex [3]. On the other hand, current problems have increased in complexity because they involve interactions between disciplines, and "to solve them it is necessary to modify how needs are managed and administered by clients and users" [3].

These conclusions are based on a review of the literature to determine the validity and effectiveness of the models for documenting the elicitation of requirements. They are strongly based on natural language, which makes their interpretation difficult and, due to ambiguities, generates re-processes in later phases of the life cycle. This requires taking the best documented practices and adding principles from logic, abstraction, and formal methods to structure a semiformal model for documenting elicitation [3].

Reviewing current practices, [6] could not find a model that would allow for formal or semiformal documentation of requirements elicitation; therefore, [3] chose to select a series of concepts that includes templates, schemas, scenarios, and diagrams. The following semiformal model is adopted:

1. Draw up the (flowchart-based) process diagram
2. Apply propositional calculus.
3. Elaborate the path diagram.
4. Complete the elicitation template
5. Generate the requirements report.

These steps are applied to an ATM network example. In this paper, we present an alternative TM model for the first three steps above.

### D. *Against Flowcharting*

Furthermore, this research has been extended to criticize all types of flowcharting. Although flowcharts have been the target of complaints regarding their value, they still have their place in the world of computing [7]. In [3]'s sample model, the flowchart is the central piece of the diagrammatic model. It is interesting to have flowcharts still play an important role in modeling, especially because they have been revived in business process model and notation (BPMN) and UML. In this paper, we suggest that static TM modeling ought to be a contemporary replacement for flowcharting.

To substantiate this extension of TM usability, TM modeling is suggested as a substitute to BPMN, which has long been a standard for modeling business processes and is currently widely used by business and technical communities [8]. In this paper, we will re-model a sample BPMN-based ordering system to contrast the two modeling methods.

### E. *Sections*

For the sake of a self-contained paper, the next section briefly concerns the model that forms the TM foundation of the theoretical development. Section 3 introduces a requirements engineering approach by [3], modeling an ATM withdrawal process. Section 4 includes a study involving an MPMN model used to describe the business process. Lastly, [3] advocates adding principles from "logic, abstraction, and formal methods to structure a semiformal model for documenting elicitation." In section 5, we show an example of TM modeling that demonstrates its suitability as a logical tool.

## II. TM MODELING

This section includes a summary of the TM model discussed in previous papers (e.g., [9-12]).

### A. *General Outlook*

The TM model provides us with an ontological representation of reality. It represents the entities that "there are" and processes them in the targeted domain utilizing one notion, which we call a *thimac*. (*thi*ng/*mac*hine). Fig. 1 shows the structure of a thimac.

A generic thimac is a gathering up of elements into unity a or synthesis of actions: create, process, release, transfer, and receive. The thimac constituents are formed from the makeup of these actions. An action is a unit of actionality. A TM diagram is called a *region* at the static-modeling level. Fig. 2 shows a picture that outlines the two levels of the TM scheme. The synthesis of actions mentioned previously is also applied to *events* (at the dynamic level—see Fig. 2).

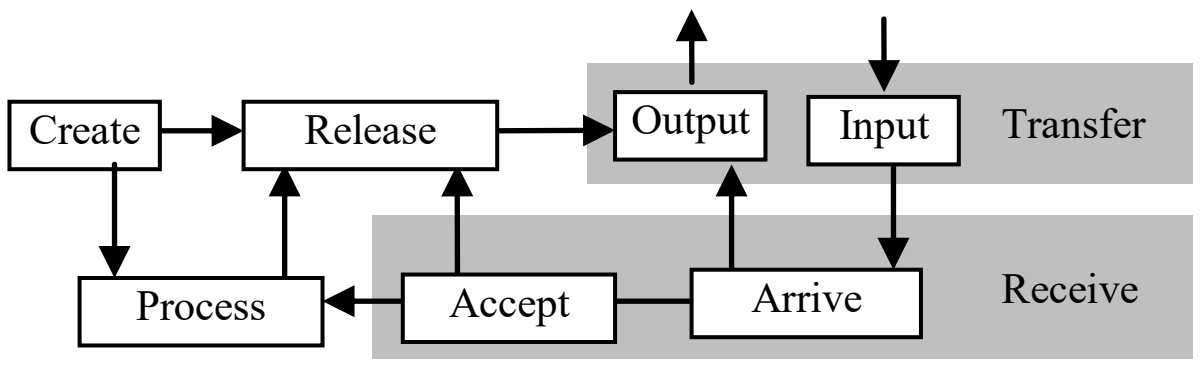

Fig. 1. Thimac

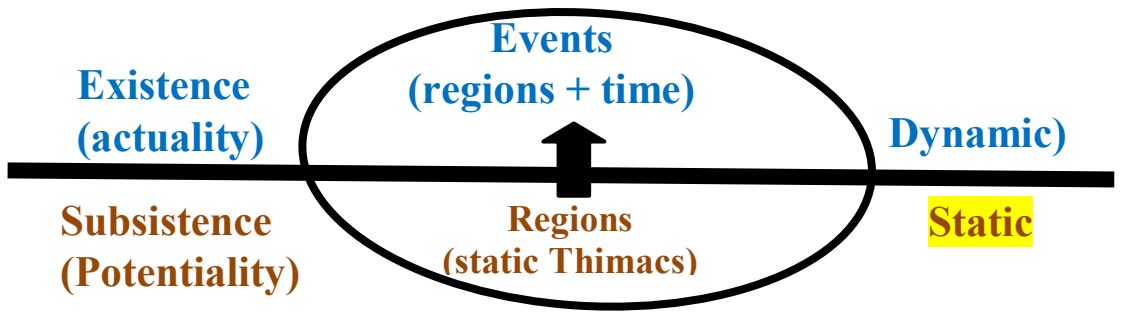

Fig. 2 Two levels of TM modeling

*B. The Thimac*

The thimac has a dual nature of a thing and a machine. It is a thing that is subjected to five actions and a machine that acts on things. A thimac's actions, shown in Fig. 1, are described as follows:

1) ***Arrive:*** A thing arrives at a thimac.
2) ***Accept:*** A thing enters a thimac. For simplification, the arriving things are assumed to be *accepted* (see Fig. 1); therefore, *arrive* and *accept* actions are combined into the ***receive*** action. Thus, the thing becomes inside the machine.
3) ***Release:*** A thing is ready for transfer outside the thimac.
4) ***Process:*** A thing is changed, handled, and examined, but no new thimac is generated.
5) ***Transfer:*** A thing crosses a boundary as input or output from a thimac.
6) ***Create***: A new thimac is registered as an ontological unit.

Additionally, the TM model includes *storage* (represented as a **cylinder** in the TM diagram) and *triggering* (denoted by a **dashed arrow**). Triggering transforms from one series of movements to another (e.g., electricity triggers heat generation).

## III. Flowchart-Based vs. TM Modeling

[3] advocates adopting the best documented practices and adding principles from logic, abstraction, and formal methods to structure a semiformal model for requirements documentation. [3] describes a sample model of an ATM case study, specifically, applied to the ATM withdrawal. Fig. 3 describes the sequence of activities included in the withdrawal module of an ATM. According to [3],

> This diagram represents the relationships between the components of the system, but it should be recognized that the system itself is immersed in another more complex system, involving the bank and its relationships with other systems. The practice of a visual representation makes it easier for stakeholders to better understand the abstract description of the problem and to interpret the requirements that the solution must satisfy.

The diagram at the bottom of Fig. 3 is called the path diagram. This graphical representation is useful to customize the requirements and acts as a bridge between the technical actors and the customer [3].

Now we develop the corresponding TM model.

*A. TM static model*

Fig. 4 Shows the TM static model that matches Fig. 3 (top). In developing Fig. 4, we tried to get as close as possible to the ATM description of Fig. 3, based on our best understanding. However, this has proven to be somewhat difficult regarding certain details, such as distinguishing between the ATM as a machine and bank functions and distinguishing between physical cards and data and certain details (e.g., the need to extract balances explicitly).

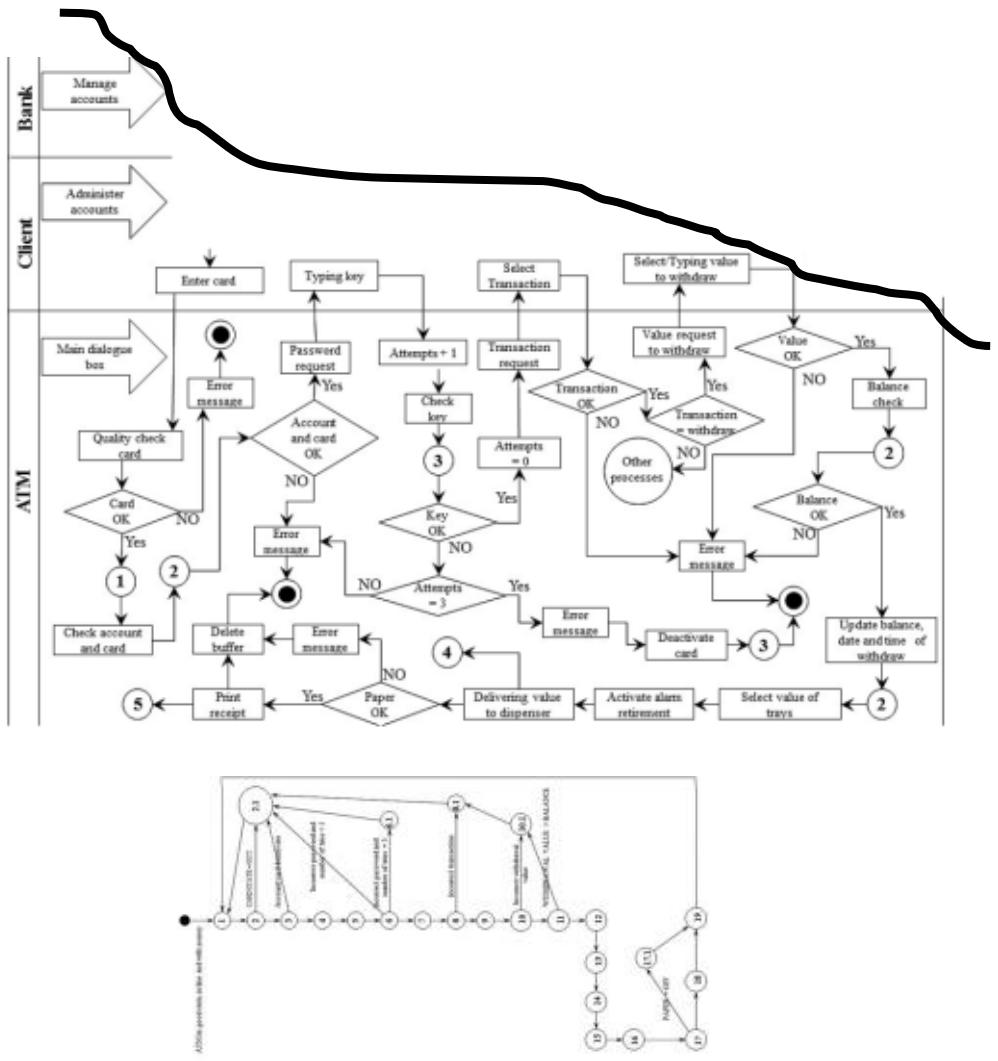

Fig. 3 Model for an ATM withdrawal module, a flowchart and path diagram (Partial, from [3])

First, we have to distinguish explicitly among the subsystems of physical cards, bank, dispenser, and the ATM system. The green numbers in Fig. 4 will be used to mark points of discussion as follows:

- The client inserts a card into the ATM machine (1).
- The card is processed (2).
- If the card is not OK (3), then an error message is output (4) and the card is rejected to flow back to the client (5). Note that the chronology of events will be specified in the dynamic model. Therefore, the flow of the card back to the client may be happen at some other point in the system (e.g., when the dispenser finishes its operations).
- If the card is OK (6), then extract the card number. Note that *Create* (at 6) indicates the appearance of the card number in its digital form for the first time in the ATM system. Before that, it was embedded on the plastic card and unknown to the system.
- The ATM system sends the card number (7) to the bank system, where it is processed (8) by searching for it in the card number/account number file (9). If it is not found (10), the bank system triggers the ATM system to send an error message to the client. Note that the processing of the card number and the file can be modeled as a loop of comparing the number with the records of the file one by one; however, for simplification, the process is represented as an input of a file and a number. This type of mixing of files and items will be used several times to reduce the diagram's complexity, especially if the search in a file for a single item is a well-known procedure.
- If the card number is found (11), then
  a) The bank system sends a message to (triggers) the ATM to continue the transaction (12).
-

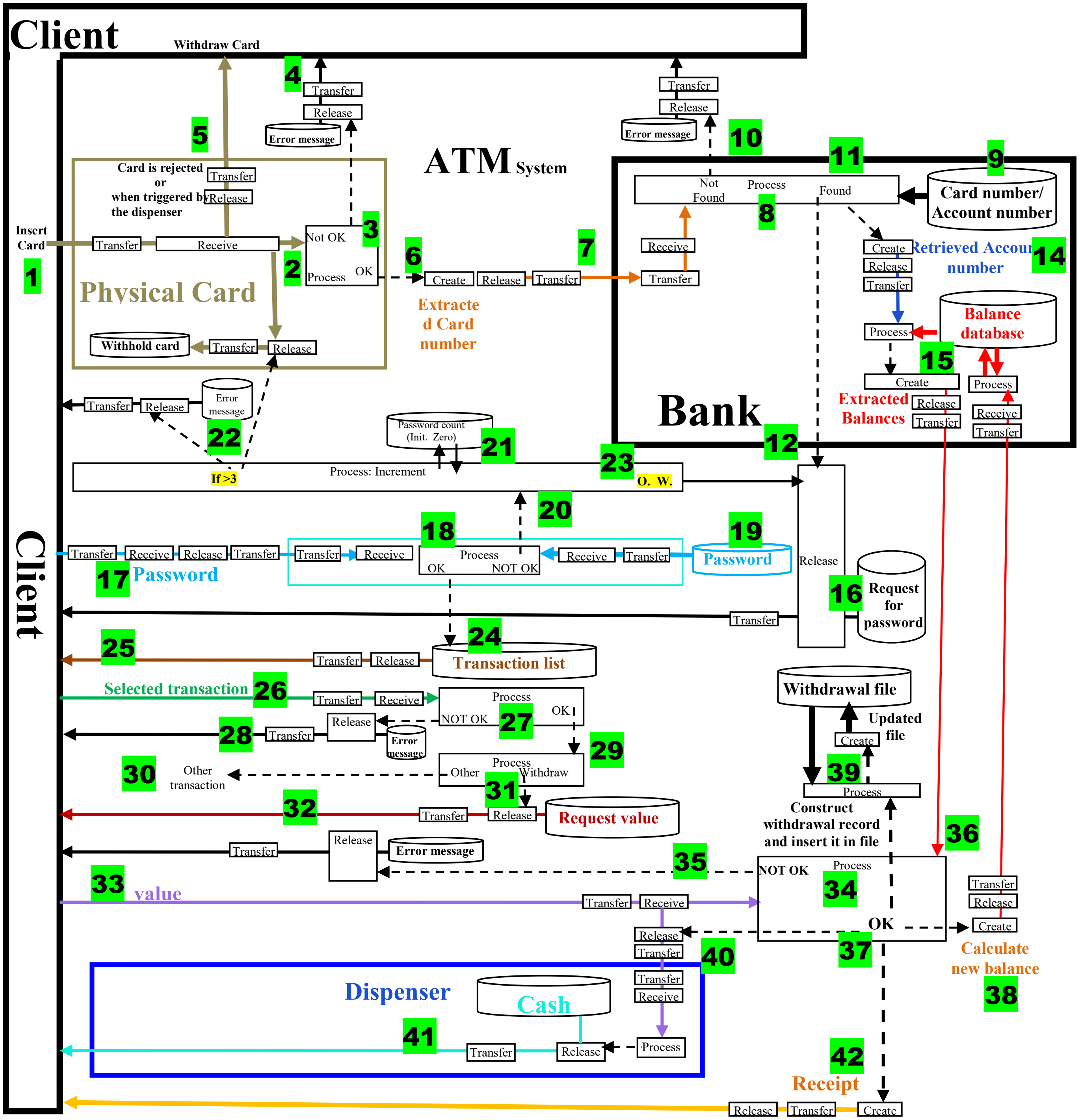

Fig. 4 TM static model of the withdrawal transaction.

b) It extracts the account number (14) and the balance (15) associated with the card number in preparation for further processing of the order.

- Accordingly, assuming reception of the OK signal from the bank (12), the ATM system sends a request for the password to the client (16). Note the request is modeled as a stored message.

- The client replies with a password (17) that is communicated to a sub-module that compares (18) the given password with the password file (19). We assume that the password file is stored in the ATM. If this is not the case, then the same communication procedure with the bank described previously can be installed here.

- A procedure to count the number of times of the password is then triggered (20). The procedure first counts the number of times, which is initially equal to zero (21), and
  a) If the number of password attempts is greater than 3, it sends an error message and confiscates the card (22).
  b) Otherwise, a new request for the password is displayed for the client (23).
- Assuming that the password is OK, then a list of transactions is sent to the client (24 and 25).
- When the transaction is received from the client (26), it is processed (27); then
  (a) If the client response is not one of the available transactions, an error message is displayed (28).
  (b) If the client response is OK (29), then if it is not a withdrawal transaction, it is directed to the appropriate system module (30). Otherwise, the withdrawal process is triggered (31).
- The withdrawal process starts with sending a request for the amount to withdraw (32). Upon receipt of the value (33), it is processed; (34) then
  (a) If the transaction is not OK (insufficient balance), an error message is sent to the client (35). Note that the current balance is provided from the bank system (36).
  (b) If the withdrawal transaction is OK (37), then the process calculates the new balance and sends it to the computer system (38), updates the withdrawal file (we assume it is on the ATM system – 39), triggers sending the value to the dispenser to deliver the cash to the client (40 and 41), and sends a receipt of the withdrawal transaction to the client (42).

### B. *Dynamic Model*

Fig. 5 shows the dynamic model; in which each event is represented by its region as follows (events are not necessarily in order):

$E_1$: A card is inserted in the ATM.
$E_2$: The card is rejected and removed from the machine (this happens in cases of card rejection or at the end of the withdrawal transaction).
$E_3$: The card is processed.
$E_4$: An error message is output because of difficulties in processing its data.
$E_5$: The card number is extracted.
$E_6$: The card number is transmitted to the bank system.
$E_7$: The account file is accessed in the bank.
$E_8$: The card number is compared with valid account numbers in the bank.
$E_9$: An error message is sent (e.g., not a valid account).
$E_{10}$: The account number is retrieved in the bank.
$E_{11}$: The database is processed to retrieve the balance.
$E_{12}$: The balanced is sent to the ATM.
$E_{13}$: The ATM system requests a password.
$E_{14}$: The client inputs the password.
$E_{15}$: The password is compared with valid passwords (assuming this file is in the ATM).
$E_{16}$: The number of password inputs is increased by 1.
$E_{17}$: The number of password inputs is greater than 3, so an error message is output and the card is confiscated
$E_{18}$: In the case of an invalid password, if the number of password inputs is less than 3, another password request is sent.
$E_{19}$: Output transaction list.
$E_{20}$: The client selects the transaction type.
$E_{21}$: The client's transaction selection is OK (e.g., on the list).
$E_{22}$: The client's transaction selection is not OK, and an error message is sent.
$E_{23}$: Process the transaction type.
$E_{24}$: If the transaction is not a withdrawal, trigger other transactions
$E_{25}$: If the transaction is a withdrawal, request an amount.
$E_{26}$: The client inputs the withdrawal amount.
$E_{27}$: The amount is compared with the balance received from the bank.
$E_{28}$: The new balance is calculated and sent to update the balance in the bank.
$E_{29}$: The withdrawal file is updated (assuming it is in the ATM).
$E_{30}$: The dispenser delivers the requested cash.
$E_{31}$: A receipt is output.
$E_{32}$: The balance is insufficient, so an error message is output.

Fig. 6 shows the event chronology diagram based on the dynamic model. A TM event is a region that is realized in time. For example, Fig. 7 shows the TM model of the event *The dispenser delivers the requested cash*. For simplicity, regions represent events in Fig. 6.

Now, it is tedious to compare the TM model with [3]'s flowchart and path diagram item by item because [3] advocates utilizing the "*best* documented practices to structure a semiformal model for documenting elicitation" (italics added). Nevertheless, the TM model is founded on a stronger foundation, in which the static and dynamic natures and chronology of events are clearly distinguished whereas the flowcharting technique is a mixture of these aspects. Additionally, the TM model is more systematic because it is based on five genetic actions whereas all English verbs may be used in flowcharting. Additionally, the TM model has a developed ontology of reality that involves regions and events.

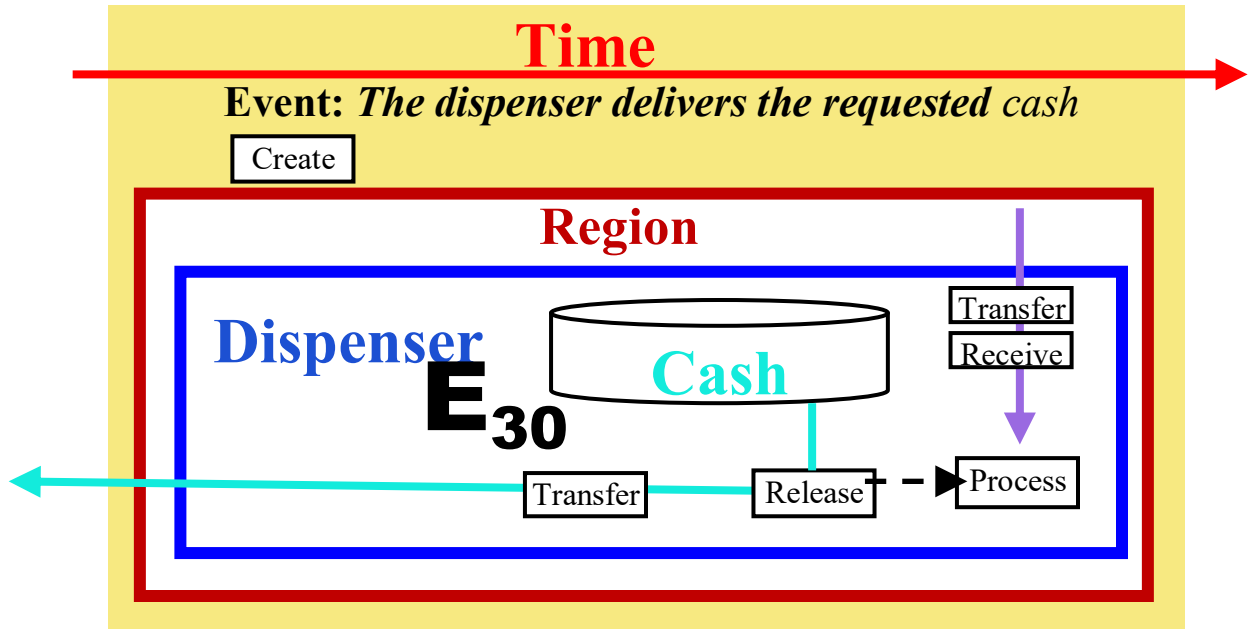

Fig. 7 The event *The dispenser delivers the requested cash.*

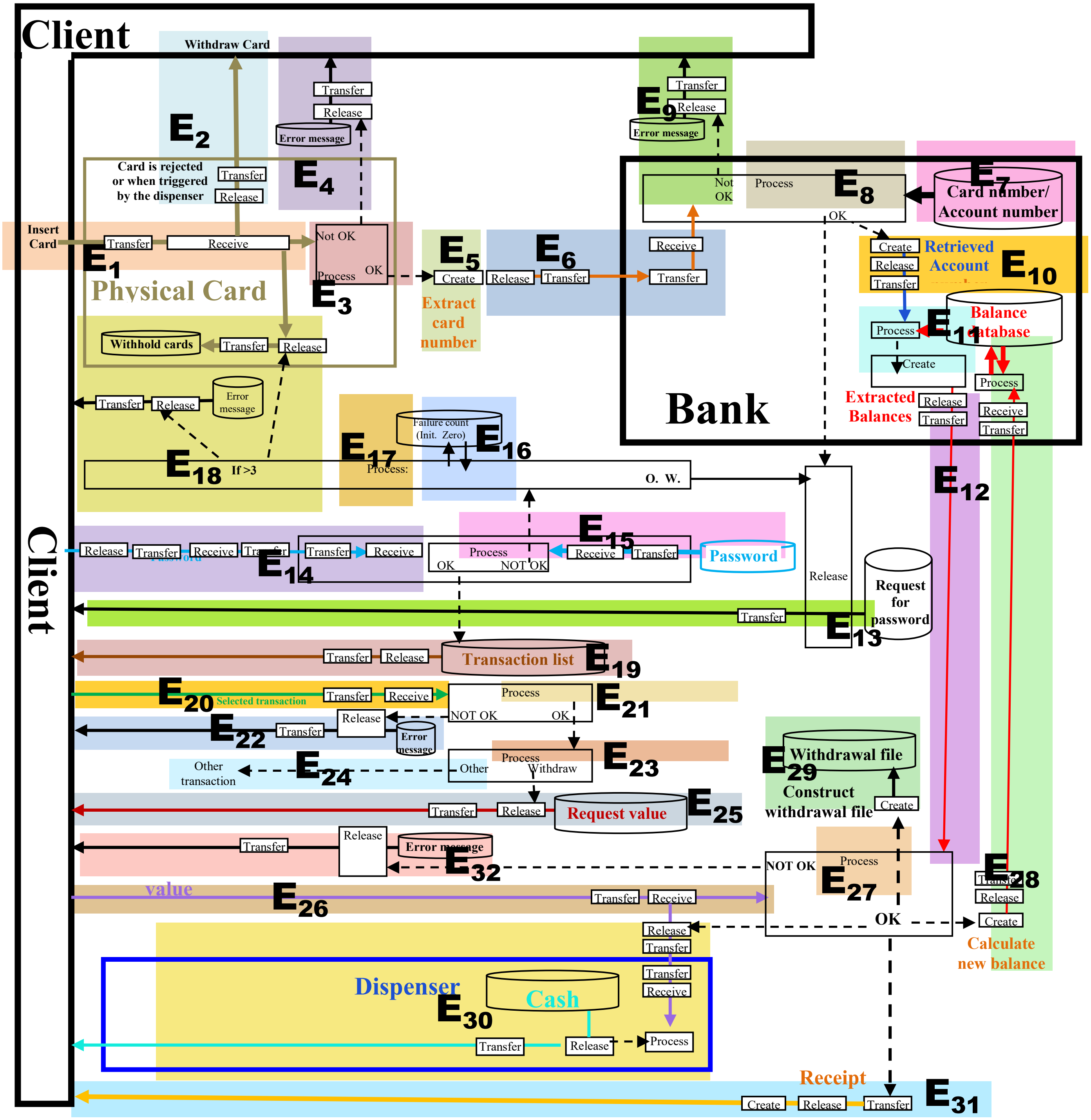

Fig. 5 The TM dynamic model.

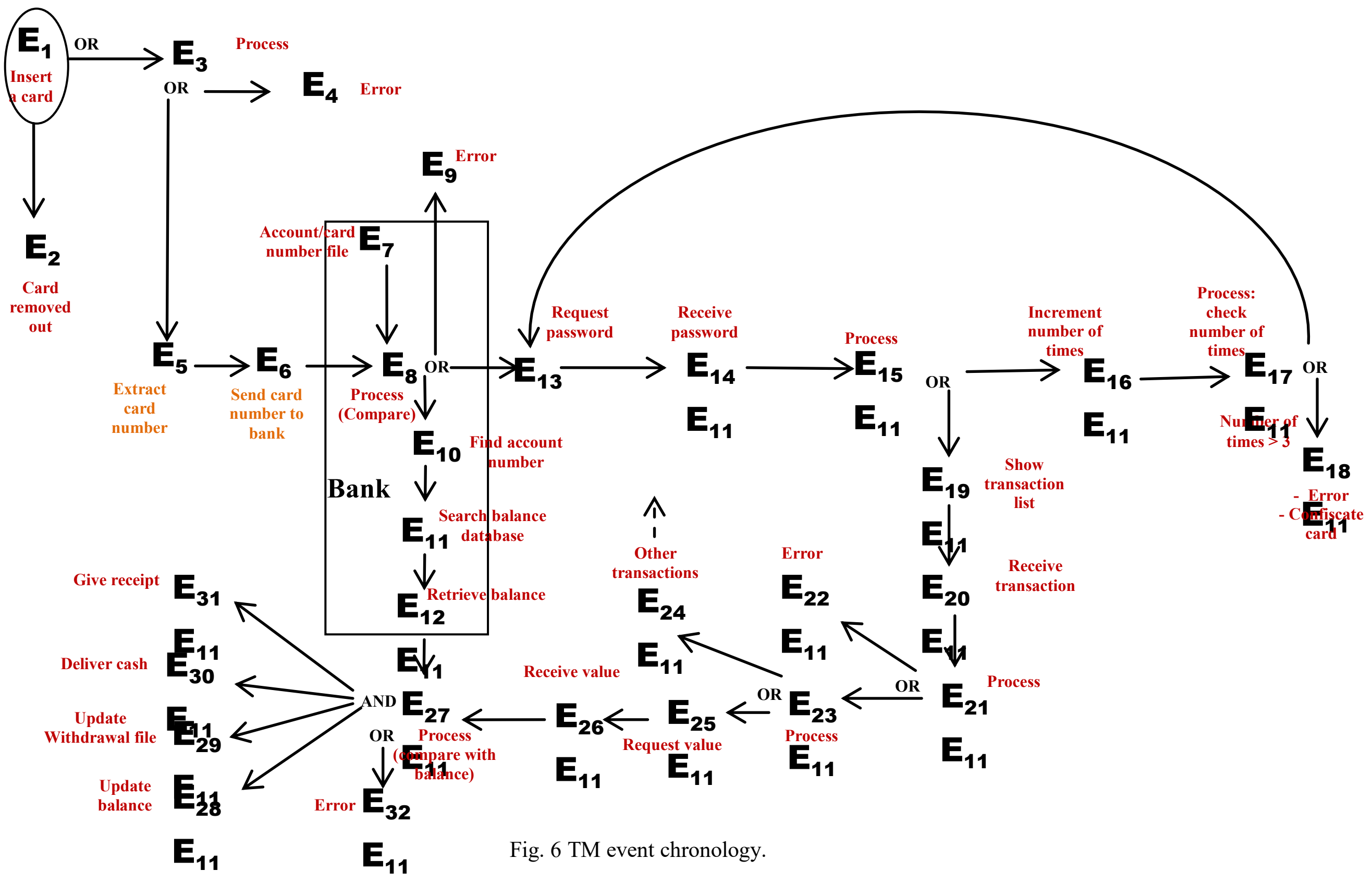

Fig. 6 TM event chronology.

## IV. Critique of Flowcharting

In [3]'s ATM withdrawal process in the previous section, the flowchart is the central piece of the diagrammatic model. We claim that the TM static model is a base for the dynamic and event chronology models and is a superior alternative to flowcharts. TM modeling seems to supersede even enriched flowcharts, such as the BPMN one. To emphasize this point, we will in this section demonstrate this claim using a recent example.

In a 2024 document [8], BPMN is used to describe the business process. According to the document, "Enterprise Architect supports a variety of ways to model processes, including the Unified Modeling Language (UML) activity diagrams, Business Process Model and Notation (BPMN). business process diagrams, and flow charts as part of the strategic diagramming set." The BPMN standard allows a modeler to document a business process, including the way the process starts, what work is performed, and how it ends. Gateways and connecting lines determine the *sequence of activities*. BPMN has long been an important standard for modeling business processes and is widely used by business and technical communities [8].

The BPMN is exemplified by modeling an ordering process, as shown in Fig. 8. We will re-model this process using TM modeling.

### A. *Static TM Model*

Fig. 9 shows the static TM model that corresponds to [8]'s (Fig. 8). We use numbers to explain the TM diagram. In Fig. 9, the ordering department receives an order (number 1 in green).

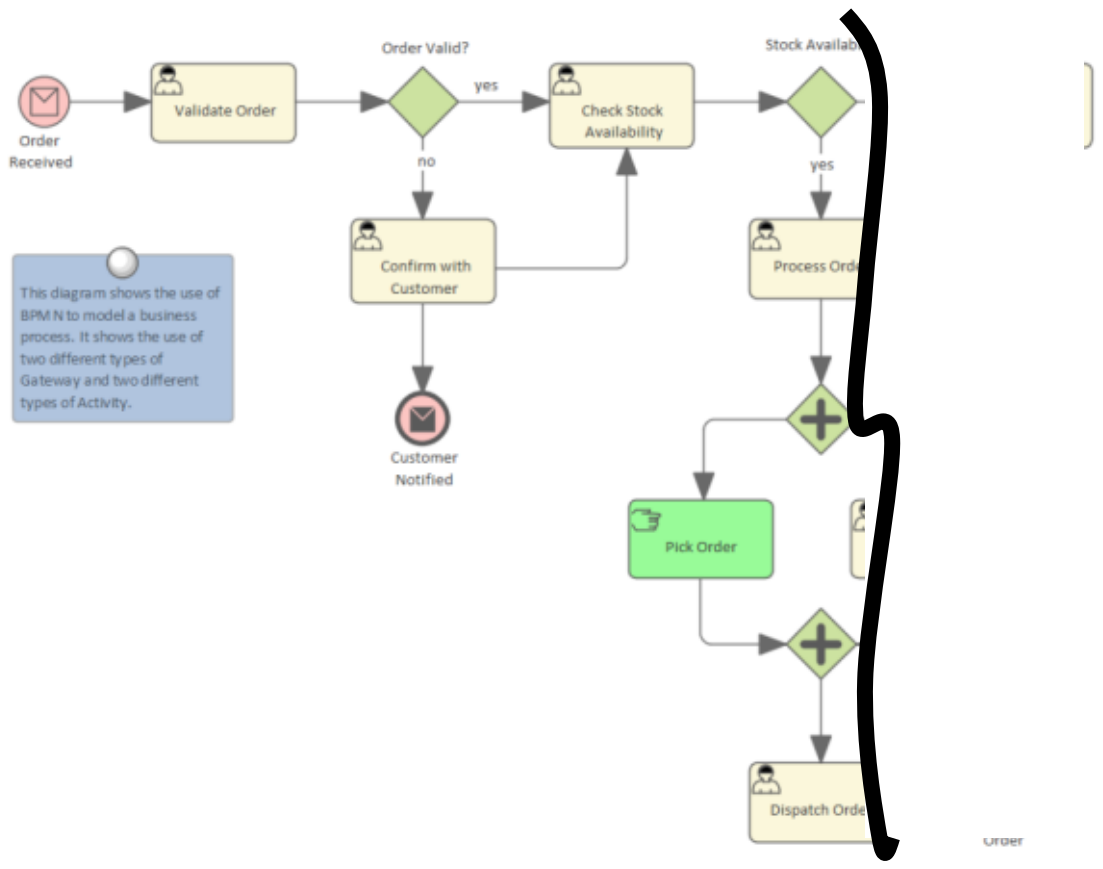

Fig. 8 Ordering process modeled in BPMN (Partial from [8SPARK]).

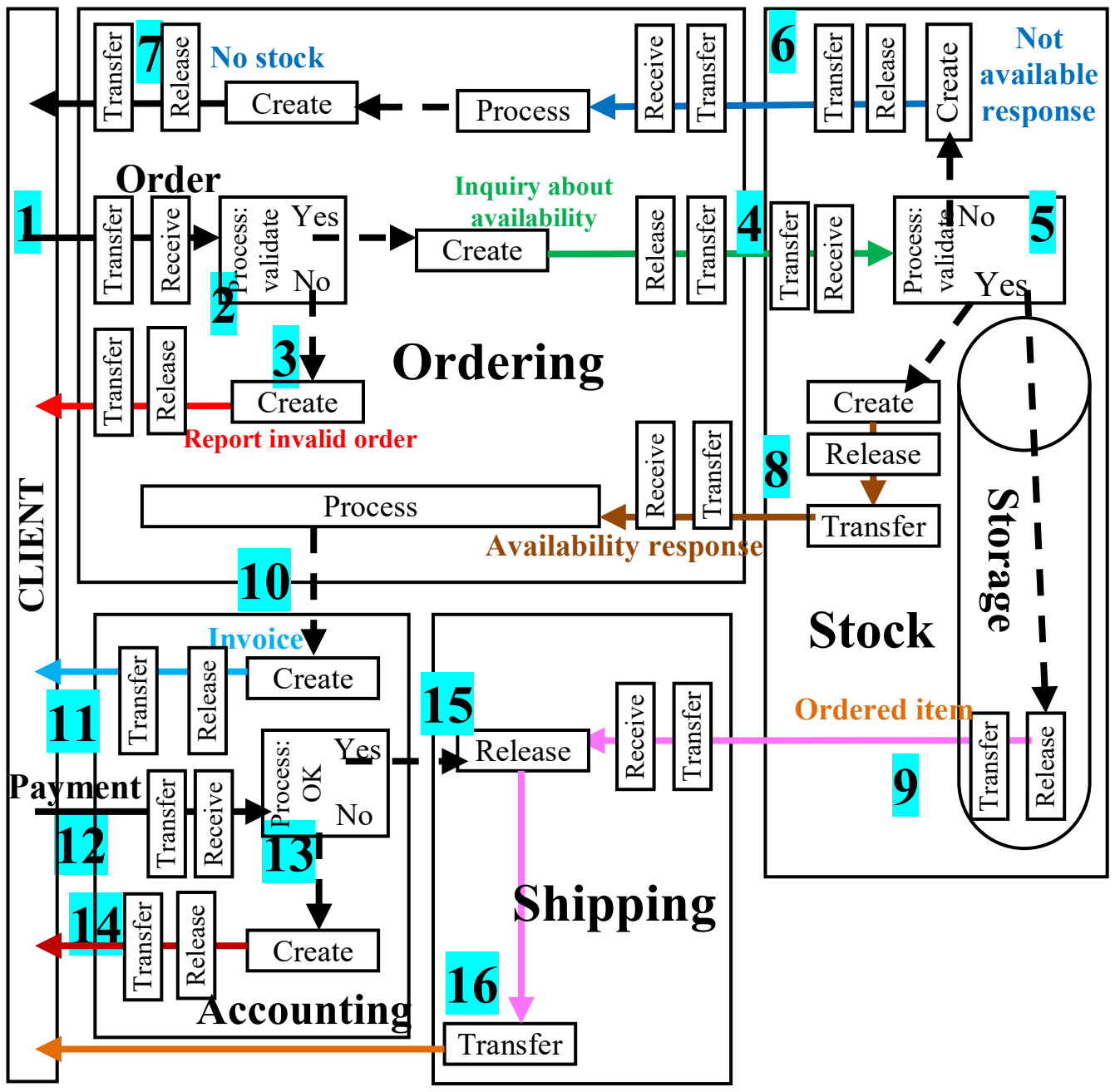

Fig. 9 TM static model of the ordering process

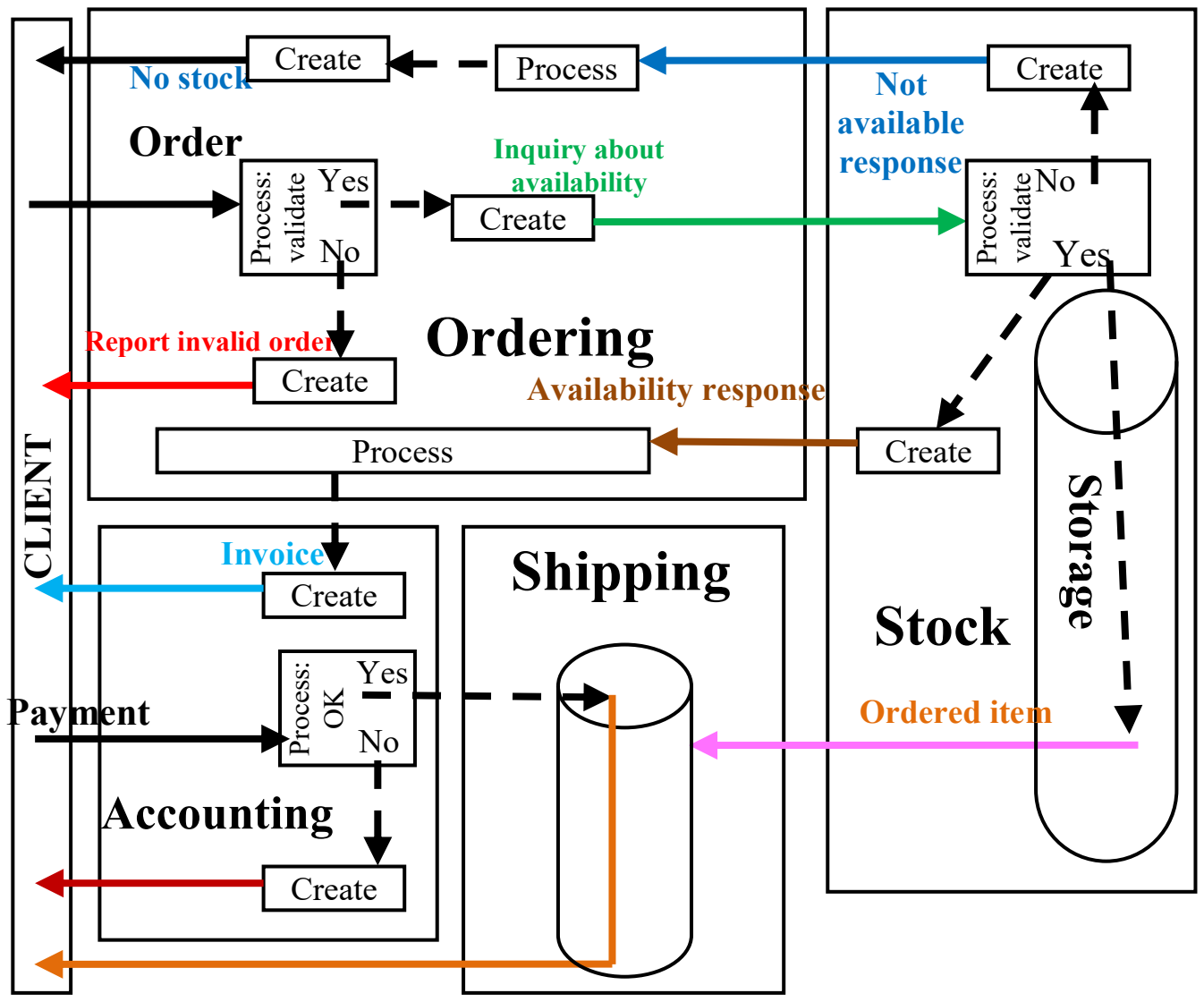

Fig. 10 Simplification of the TM static model of the ordering process

The order is processed (2), and then according to the result of processing,

- If the order is invalid, a report is sent to the client (3).
- If the order is valid, then an inquiry is sent to the stock department inquiring about the ordered item's availability (4).

In the stock department, the inquiry about availability is processed.

- If the ordered item is not in stock, a notice is sent to the ordering department (6). In this case, a notification is sent to the client about the ordered item's unavailability (7).
- If the ordered item in available, then
  - (a) A notification is sent to the ordering department to continue processing the order (8).
  - (b) The ordered item is sent to the shipping department (9).
- The ordering department triggers the accounting department to send an invoice (10 and 11), the client sends the payment (12), and the payment is processed (13).
- Accordingly,
  - (a) If the payment is not OK, a notification is sent to the client (14).
  - (b) If the payment is OK, then the shipping department is triggered (15) to deliver the ordered item to the client (16).

The TM static model can be simplified by removing the release, transfer, and receive actions (See Fig. 10) under the assumption that the arrows indicate the direction of the flow.

### B. *TM Dynamic Model*

Even though the term *event* in [8] is used 27 times, it is not defined. The OMG's BPMN 2.0 specification defines an event as "an event is something that 'happens' during the course of a process. These events affect the flow of the process and usually have a cause or an impact and in general require or allow for a reaction" [13]. According to [13], this specification of the term event is general enough to cover many things in a process (e.g., the start of an *activity*, the end of an activity, a document's change of state, a message that arrives, and a message that is sent). According to BPMN 2.0 specification, an activity represents "*work* that a company or organization performs using business processes" [14] (italics added).

Instead of these general BPMN terms, the TM model defines actions as create, process, release, transfer, or receive, which form regions of thimacs. An event is *an action in time*. Accordingly, we can construct the events in Fig. 9 as shown in Fig. 11, which includes the following events:

$E_1$: A client sends an order.
$E_2$: The order is processed.
$E_3$: The order is unacceptable and therefore returned to the client.
$E_4$: The order is accepted, and a check is conducted to verify that the ordered item is in stock.
$E_5$: In the stock department, the order is processed.
$E_6$: The ordered item is not available, so a notification is sent back to the ordering department.
$E_7$: A notification of unavailability is sent to the client.
$E_8$: The ordered item is available, and a notification is sent to the ordering department to continue processing the order.
$E_9$: The ordered item is available, and the item is delivered to the shipping department.

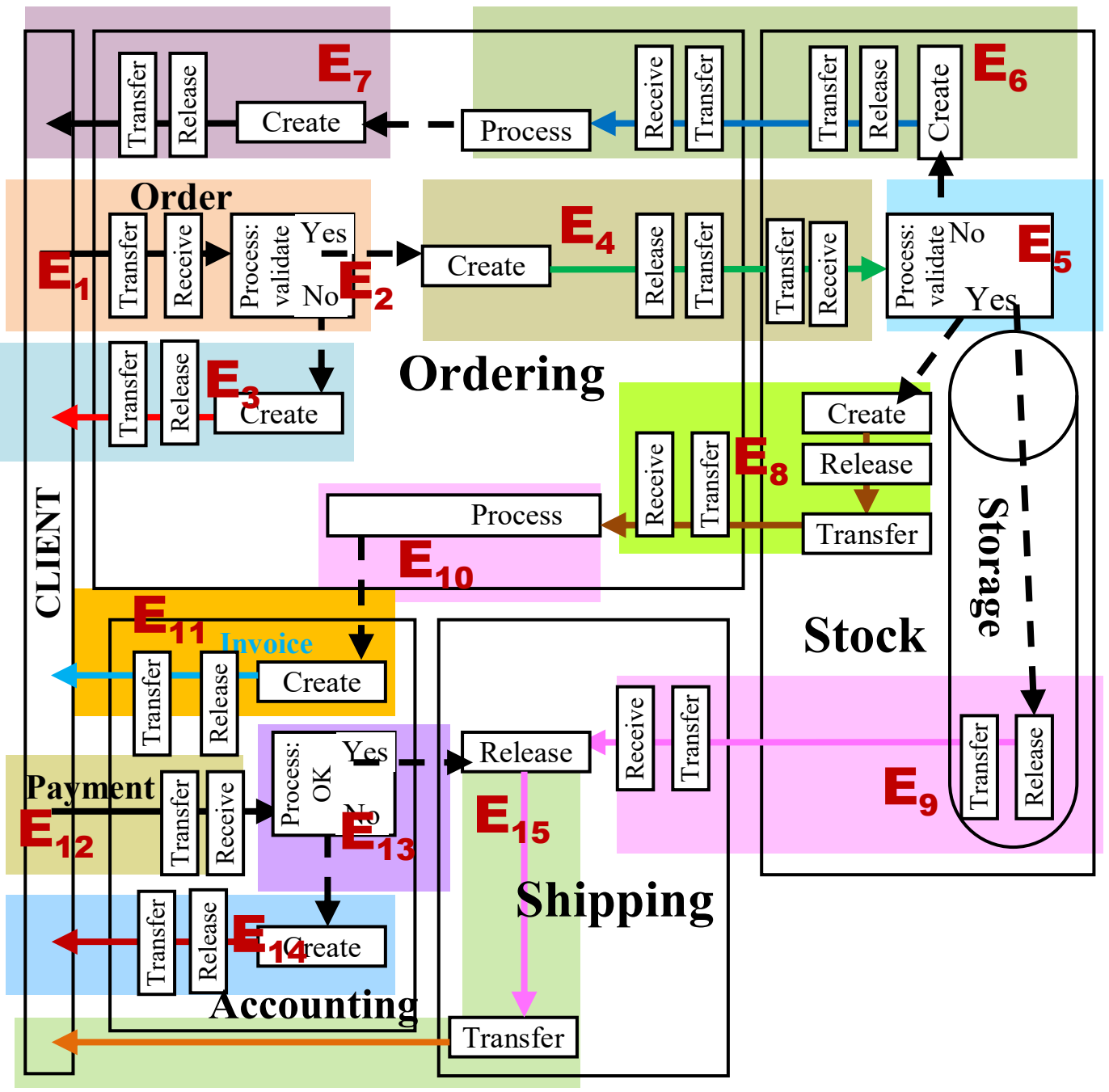

Fig. 11 The TM dynamic model of the ordering process.

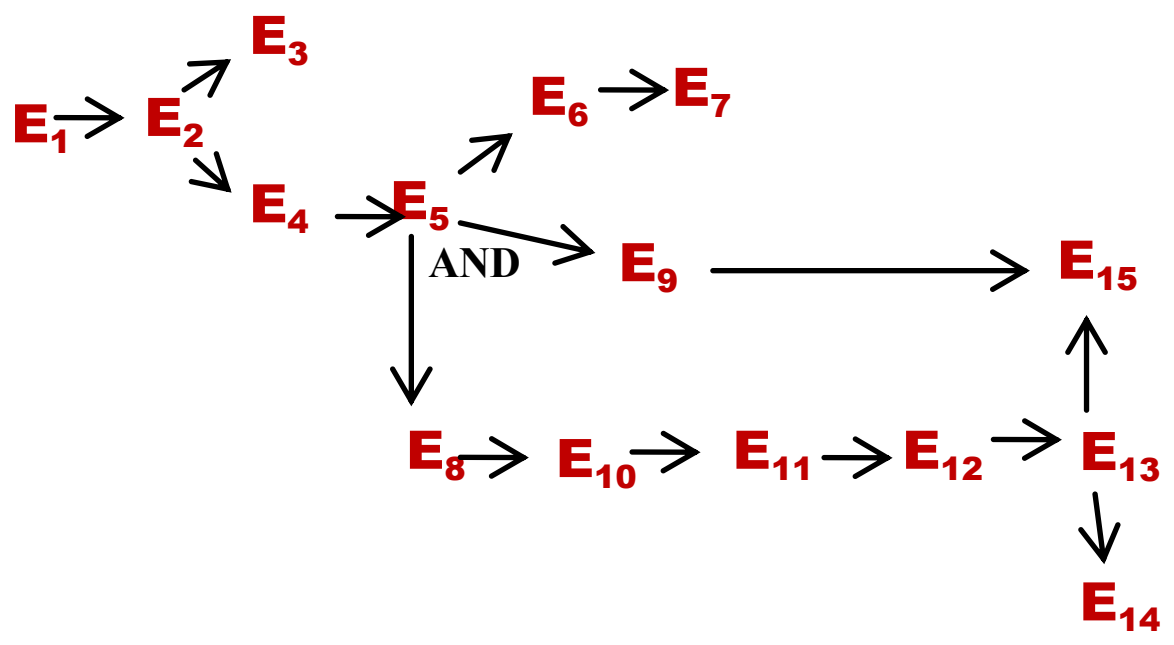

Fig. 12 The chronology of events.

$E_{10}$: The ordering department triggers the accounts department to send an invoice to the client.
$E_{11}$: The accounts department sends an invoice to the client.
$E_{12}$: The client sends a payment.
$E_{13}$: The payment is processed.
$E_{14}$: The payment is not OK, and a notification is sent to the client.
$E_{15}$: The payment is OK.
$E_{16}$: The shipping department delivers the item to the client.

Fig. 12 shows the chronology of events in Fig. 11.

Contrasting the TM and BPMN methodologies reveals that the TM approach is preferable, for the static and dynamic features are separated at two levels. The events relationships can be specified in a third level.

## V. TM FOR LOGIC

Reference [3] advocates taking the best documented practices and adding principles from "logic, abstraction, and formal methods to structure a semi-formal model for documenting elicitation." According to [5], since the 1970s, researchers have investigated the use of formal methods that are based on logic, set theory, or Petri Nets (e.g., to derive formal specifications for software systems). However, this use has not advanced as much in software engineering as in other engineering fields.

We claim that TM modeling is a suitable tool to enrich logical analysis. Initial indications of this are shown in [15], titled *Toward Conceptual Modeling for Propositional Logic: Propositions as Events*. In this section, we show that the TM approach can help in directing logical inferences.

### A. *Logical Arguments*

An argument is an assertion that a set of statements, called the premises, yields another statement, called the conclusion. An argument is valid if and only if the conjunction of the premises implies the conclusion [16]. [16] takes the following famous example from Lewis Carroll [17].

Prove that it is a valid argument.

1. All the dated letters in this room are written on blue paper.
2. None of them are in black ink, except those that are written in the third person.
3. I have not filed any of those that I can read.
4. None of those that are written on one sheet are undated.
5. All of those that are not crossed out are in black ink.
6. All of those that are written by Brown begin with "Dear Sir."
7. All of those that are written on blue paper are filed.
8. None of those that are written on more than one sheet are crossed out.
9. None of those that begin with "Dear Sir" are written in the third person.

Therefore, I cannot read any of Brown's letters.

According to [16], with a little patience, these propositions can be reduced to pairs of *eliminands*, which cancel each other out, leaving two *retinands*, "letters written by Brown" and "letters that I cannot read," comprising the conclusion "I cannot read any of Brown's letters." [16] presents the argument in propositional logic, as shown in Fig. 13.

Let
**p be "the letter is dated,"**
**q be "the letter is written on blue paper,"**
**r be "the letter is written in black ink,"**
**s be "the letter is written in the third person,"**
**t be "the letter is filed,"**
**u be "I can read the letter,"**
**v be "the letter is written on one sheet,"**
**w be "the letter is crossed out,"**
**x be "the letter is written by Brown,"**
**y "the letter begins with 'Dear Sir' "**

1. $p \to q$
2. $\neg s \to \neg r$
3. $u \to \neg t$
4. $v \to p$
5. $\neg w \to r$
6. $x \to y$
7. $q \to t$
8. $\neg v \to \neg w$
9. $y \to \neg s$

Therefore $x \to \neg u$

Fig. 13 The given premises and the conclusion (From [16]).

The reader may try to understand a sample solution that is given in Fig. 14 [18] (Good luck!).

Fig. 15 models the given premises and conclusion in TM. Each proposition that has negative and positive versions is represented by two thimacs attached to each other. The positive thimac is in color (e.g., r), and the negative thimac has a dotted perimeter (e.g., ¬r). TM modeling of negative proposition has been discussed in [15].

In Fig. 15, simplification is applied in certain cases, such as representing the *letter* as *(by) brown* subthimac. Each triggering corresponds to an implication. In Fig. 15,

Circle 1 corresponds to **x→y**
Circle 2 corresponds to **y→¬s**
Circle 3 corresponds to **¬s→¬r**
Circle 4 corresponds to **¬w→r**
Circle 5 corresponds to **¬v→¬w**
Circle 6 corresponds to **v→p**
Circle 7 corresponds to **p→q**
Circle 8 corresponds to **q→t**
Circle 9 corresponds to **u→¬t**
Circle 10 corresponds to **x→¬u**

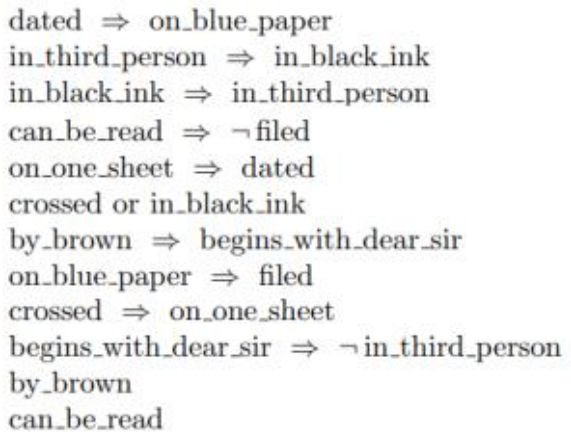
dated ⇒ on_blue_paper
in_third_person ⇒ in_black_ink
in_black_ink ⇒ in_third_person
can_be_read ⇒ ¬filed
on_one_sheet ⇒ dated
crossed or in_black_ink
by_brown ⇒ begins_with_dear_sir
on_blue_paper ⇒ filed
crossed ⇒ on_one_sheet
begins_with_dear_sir ⇒ ¬in_third_person
by_brown
can_be_read

Fig. 14 Sample solution (From [18]).

Fig. 16 shows the inference path with thick orange arrows. The difficulty in this inference path clearly arises in crossing from

(a) **¬r to w and w to v**: {¬w→r, ¬r} implies w, and {¬v→¬w, w} implies v.
(b) **t to x→¬w**: here, u→¬t, t} implies the required result, x→¬w,

as shown in Fig. 17.

In this process, the TM model forms the map for the path to the solution. The TM modeling may present a semantic foundation for logic, as demonstrated for propositional logic in [15].

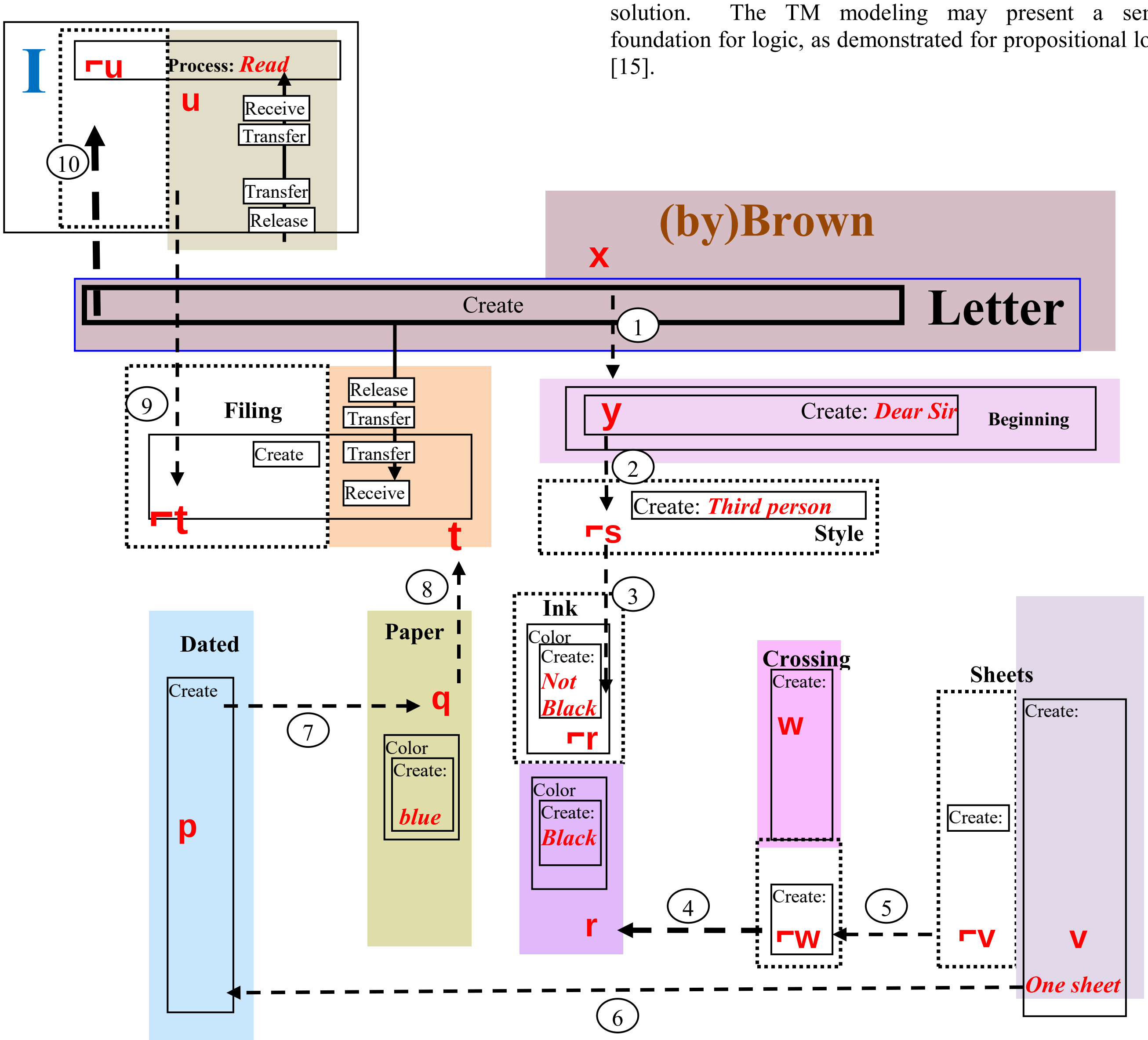

Fig. 15 The dynamic model of the given logic example.

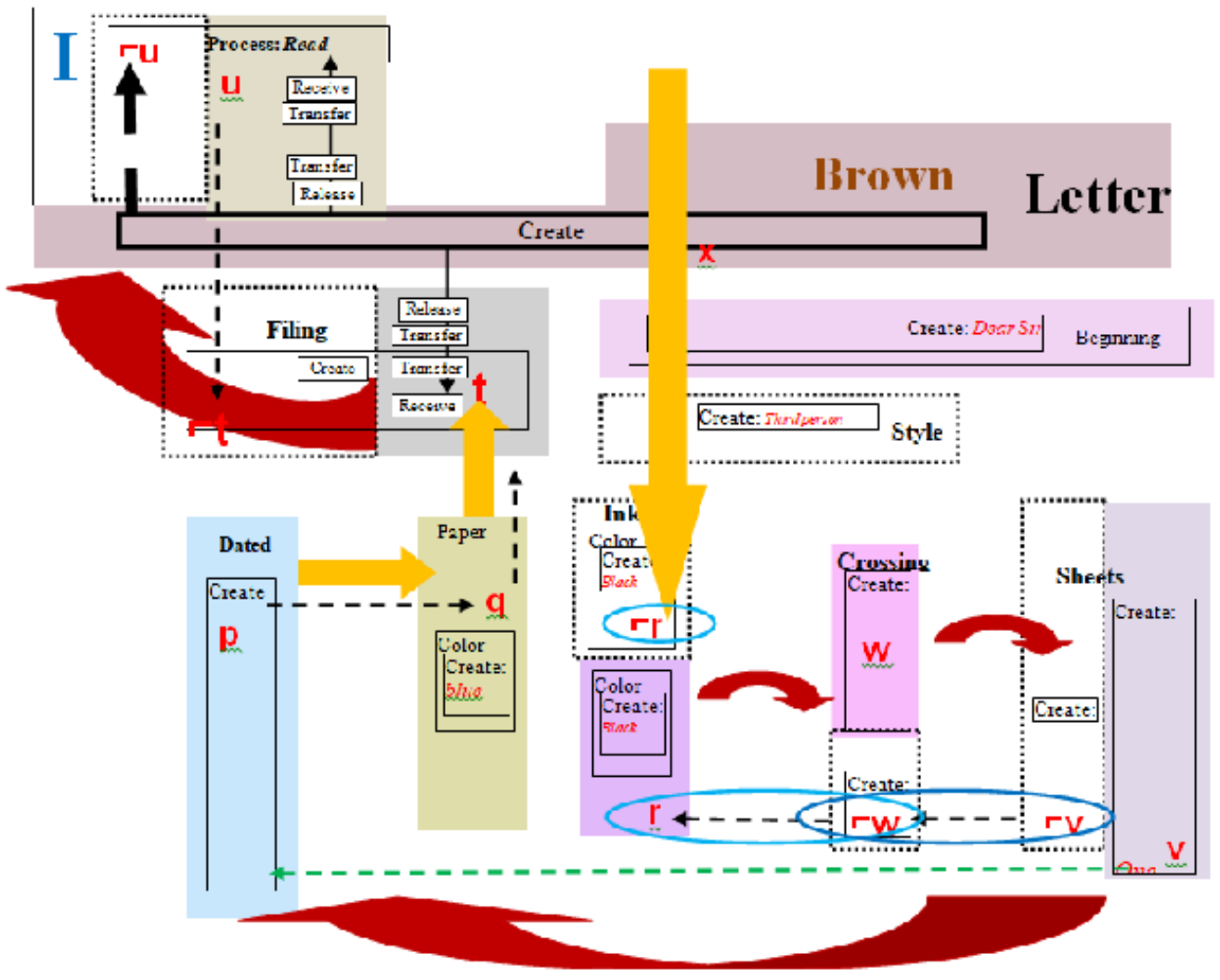

Fig. 16 Inference path.

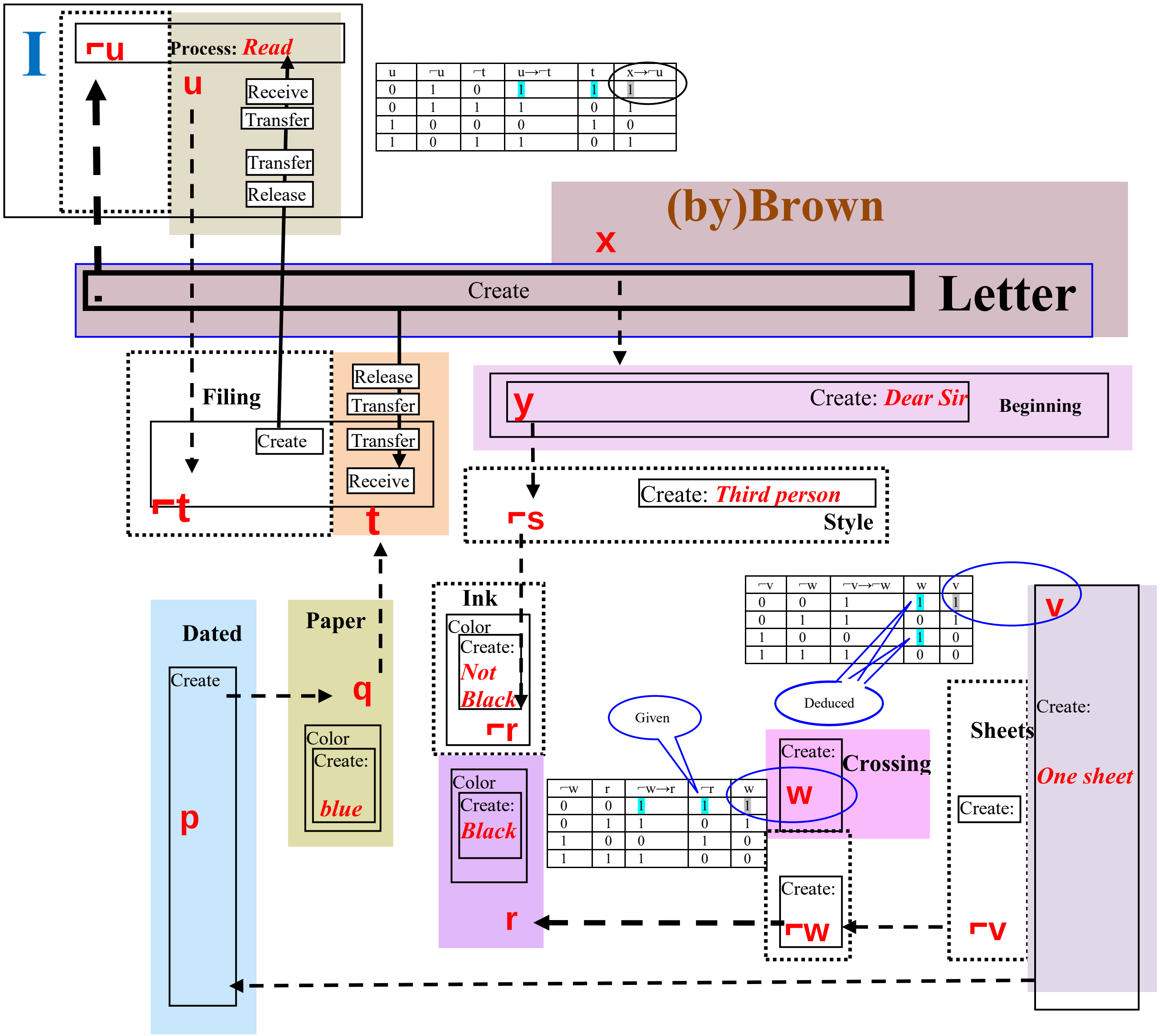

| u | ⌐u | ⌐t | u→⌐t | t | x→⌐u |
|---|---|---|---|---|---|
| 0 | 1 | 0 | 1 | 1 | 1 |
| 0 | 1 | 1 | 1 | 0 | 1 |
| 1 | 0 | 0 | 0 | 1 | 0 |
| 1 | 0 | 1 | 1 | 0 | 1 |

| ⌐v | ⌐w | ⌐v→⌐w | w | v |
|---|---|---|---|---|
| 0 | 0 | 1 | 1 | 1 |
| 0 | 1 | 1 | 0 | 1 |
| 1 | 0 | 0 | 1 | 0 |
| 1 | 1 | 1 | 0 | 0 |

| ⌐w | r | ⌐w→r | ⌐r | w |
|---|---|---|---|---|
| 0 | 0 | 1 | 1 | 1 |
| 0 | 1 | 1 | 0 | 1 |
| 1 | 0 | 0 | 1 | 0 |
| 1 | 1 | 1 | 0 | 0 |

Fig. 17 The dynamic model with truth tables.

## VI. CONCLUSION

In conclusion, we claim that the TM model makes many types of flowcharting-based modeling obsolete techniques. The TM model includes an extension of the basic static model to specify a system's dynamic behaviors. The advantages of the resulting diagrams include an events chronology description based on two-level ontology.

At this stage, such an informal contrast between classical flowcharting and the new approach is a reasonable way to persuade workers in this area of the new approach's potential. Substantial research is needed, but the demonstrated outcomes, through contrasting classical flowcharting and the new approach, appear adequate to show the advantages, including a more systematic diagrammatic description (in comparison with flowcharts).